\documentstyle[11pt,newpasp,twoside,epsf,psfig]{article}
\def\gs{\mathrel{\raise0.35ex\hbox{$\scriptstyle >$}\kern-0.6em
\lower0.40ex\hbox{{$\scriptstyle \sim$}}}}
\def\ls{\mathrel{\raise0.35ex\hbox{$\scriptstyle <$}\kern-0.6em
\lower0.40ex\hbox{{$\scriptstyle \sim$}}}}

\reversemarginpar

\begin{document}

\title{A Submm View of the Universe: Clues to the Formation of
Massive Galaxies}
 \author{Ian Smail}
\affil{Institute for Computational Cosmology, University of Durham,
Durham DH1 3LE, UK}
\author{S.C.\ Chapman, A.W.\ Blain}
\affil{California Institute of Technology, Pasadena, CA91125, USA}
\author{R.J.\ Ivison}
\affil{Astronomy Technology Centre, Royal Observatory, Blackford
Hill, Edinburgh EH9 3HJ, UK}

\begin{abstract}
We discuss recent advances in the study of dusty, massive galaxies
at $z>1$ arising from the first extensive spectroscopic surveys.
\end{abstract}

\section{Introduction}

There is a growing evidence for the existence of a population of
massive galaxies at comparatively early epochs in the Universe, $z\gs
1$--2 (e.g.\ Franx et al.\ 2003; Glazebrook et al.\ 2004). As extreme
examples of the process of galaxy formation these systems provide a
strong test of theoretical models and an excellent opportunity to
observe the interplay of the physical processes involved in galaxy
formation.  The first question that this population poses is: When and
how were these massive galaxies formed?  Are they assembled on a
timescale comparable to their star-formation in a single massive burst,
or from a more prolonged build-up from sub-components which had already
undergone substantial star formation?

Observations from the local Universe provide some information on the
formation history of massive galaxies.  The most luminous, and by
implication most massive, galaxies in the local Universe are the giant
ellipticals (gE's) which frequently reside in high-density regions such
as clusters.  The stellar population of these systems are dominated by
old, metal-rich stars suggesting that the bulk of their star-formation
activity occurred at high redshifts and hinting at a connection with
the massive galaxies seen at high-redshifts, including the passive
component of the Extremely Red Object (ERO) population.  Quantitative
constraints can be placed on the formation redshift from the
identification of luminous early-type galaxies in clusters out to
$z\sim 1.2$, as well as the homogeneity of the colours of these bright
galaxies, which imply their stars were formed at $z>2$--3 (Ellis et
al.\ 1997; Blakeslee et al.\ 2003).

The gE's have luminosities of $\geq 1$--10L$^\ast_V$
(10$^{10}$--10$^{11}$L$_\odot$) and space densities of $\sim
10^{-4}$--$10^{-5}$\,Mpc$^{-3}$ respectively.  The central 1-Mpc of a
rich cluster can easily contain $2\times 10^{12}$L$_\odot$ of stars in
the $\geq$L$^\ast$ gE population, as well as a substantial amount of
metals in the intracluster medium ($7\times 10^{10}$M$_\odot$ of Fe,
Renzini 1997).  Further insights into the mode of their assembly comes
from modeling the variation in colour of the elliptical population with
luminosity (the colour-magnitude relation) and the internal colour
gradients within individual elliptical galaxies.  Both of these
observational trends can be reproduced in models where the metallicity
of the stellar populations in gE's is regulated by the on-set of a
superwind in a single, massive starburst (Kodama \& Arimoto 1998;
Tamura et al.\ 2000).  The presence of a substantial mass of metals in
the intracluster medium (and little evolution in this fraction out to
$z\sim 0.8$) also suggests significant outflows of material from the
cluster population at $z>1.5$ (transporting energy as well as the
metals) and lends credence to the existence of intense, high-density
and wind-generating starbursts during the formation of these massive
galaxies.

%
%
\begin{figure}
\centerline{
\vbox{\psfig{file=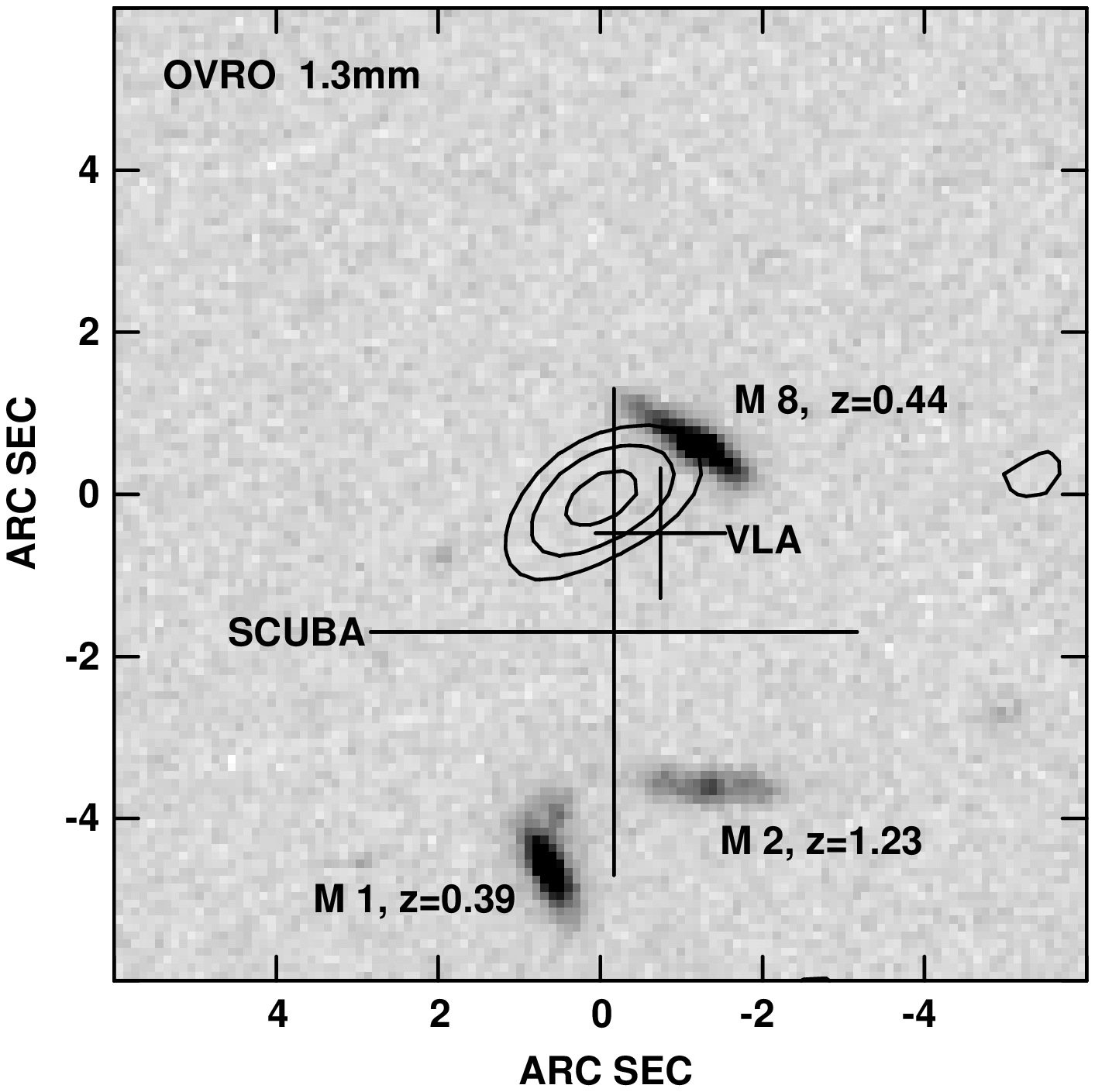,width=1.32in,angle=0}
\psfig{file=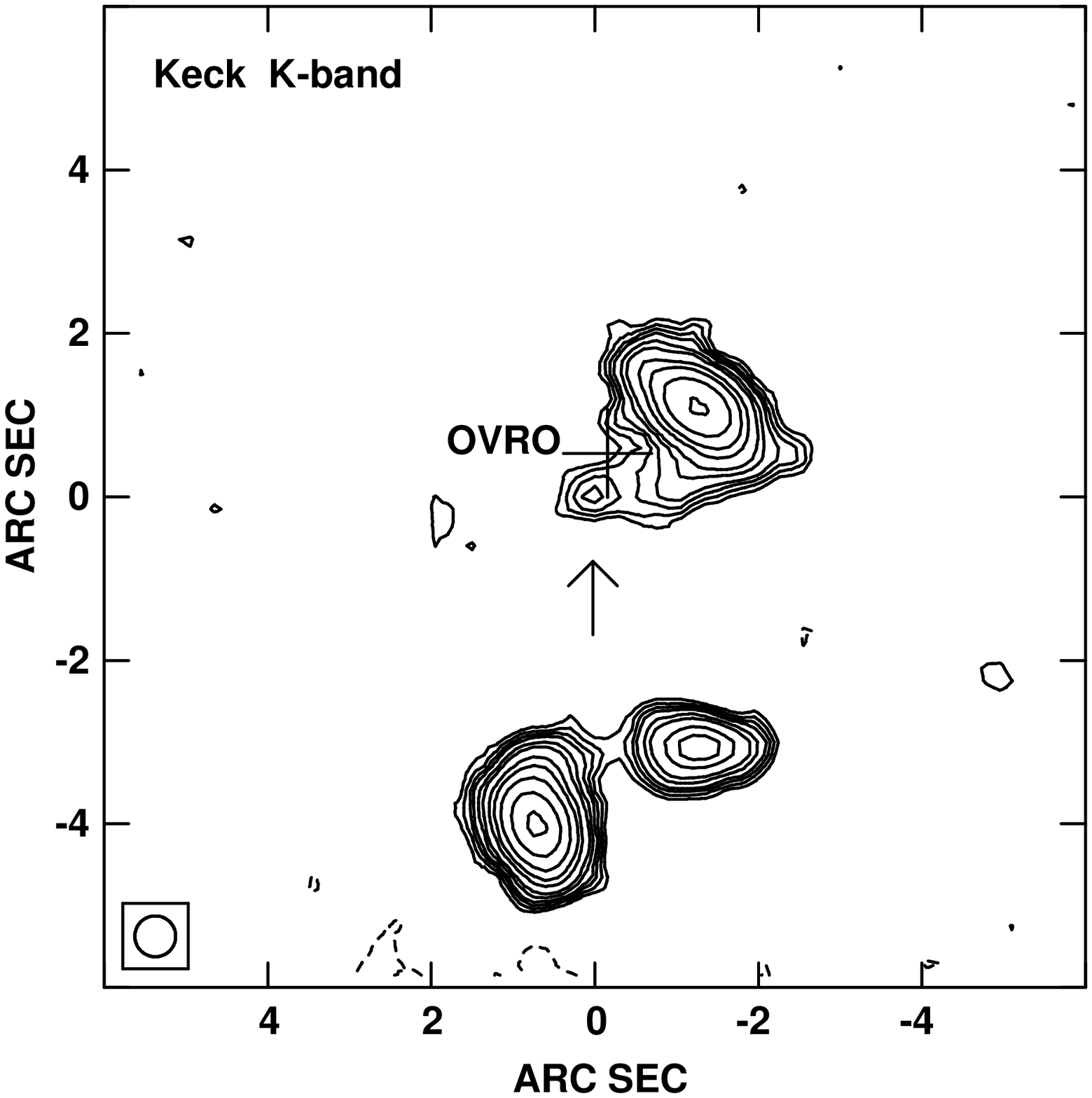,width=1.32in,angle=0}} ~~~
\psfig{file=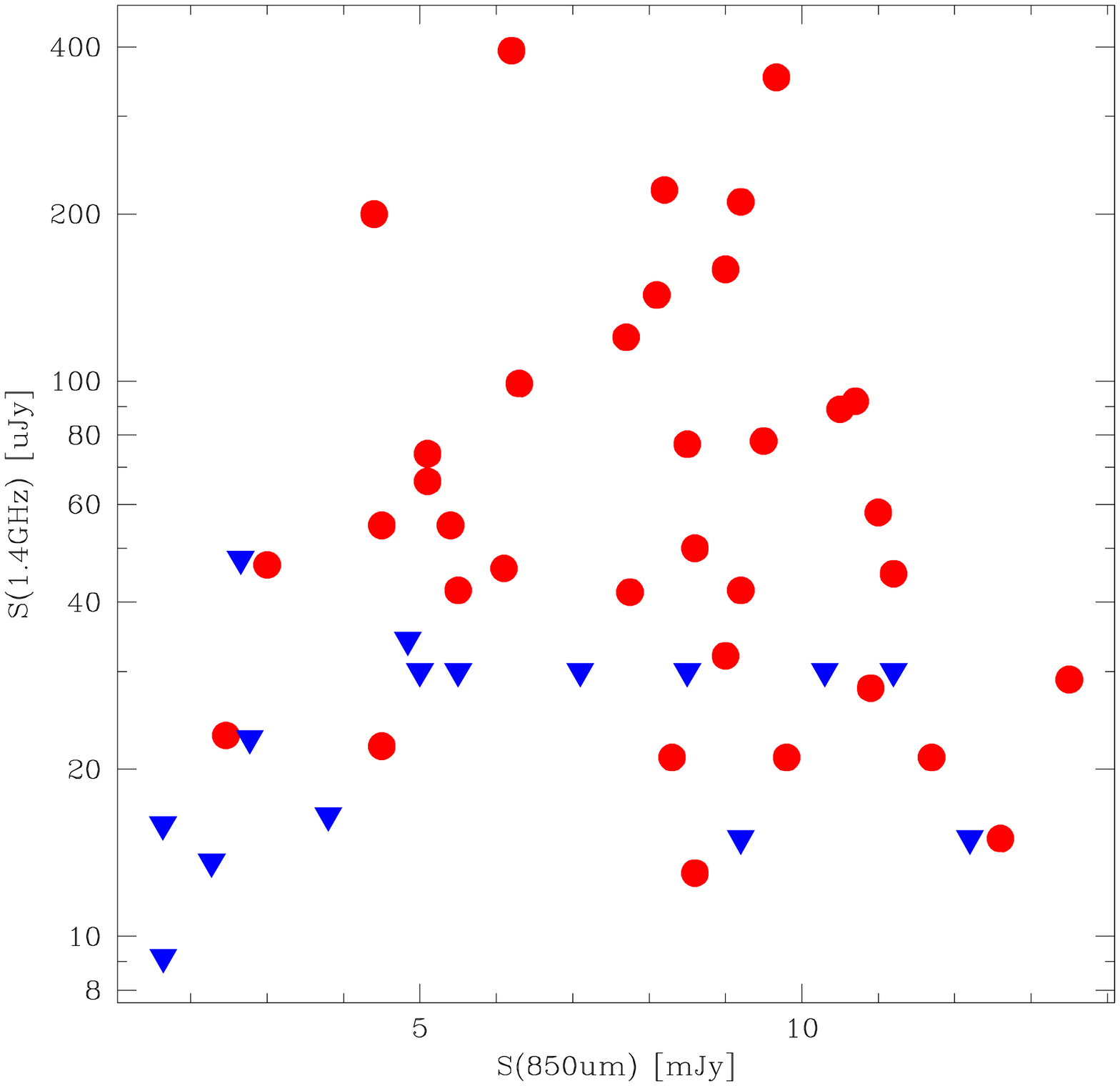,width=3.0in,angle=0}}
\caption{\small
[left] Different views of an SMG from the SCUBA Lens Survey (from
Frayer et al.\ 2000).  The top panel shows a deep {\it HST} F814W
image, the large cross indicates the position and size of the SCUBA
error-circle, which encompasses several plausible faint counterparts.
The smaller cross shows the 1.4-GHz VLA position and error-box
indicating that the source is close to, but not coincident with the
galaxy M8.  The contour plot is a deep millimetre continuum map from
the OVRO interferometer which confirms that the submm source lies in an
apparently blank area of sky.  The lower panel is a contour plot of a
very deep Keck K-band image of this field which detects a red K$\sim$22.5
counterpart, M12, to the radio/mm-identified SMG.  [right] The
radio fluxes (or upper limits) as a function of submm flux for 
SMGs from the 8-mJy survey (Ivison et al.\ 2002) and the SCUBA Lens
Survey (Smail et al.\ 2002).  This demonstrates that  60--70\%
of SMGs brighter than $\sim 5$mJy have radio counterparts in deep
1.4-GHz maps ($\sigma_{1.4}\sim 5$--8$\mu$Jy).}
\end{figure}

What other observational tests can distinguish between the possible
formation scenarios of massive galaxies? If the activity required to
produce the population of massive galaxies in a rich cluster occurred
over a short period of time, say $<1$\,Gyrs (or between $z=2$--3), this
would need an average star formation rate (SFR) of
$>2000$\,M$_\odot$\,yr$^{-1}$ in a volume of roughly 1000\,Mpc$^{-3}$.
The visibility of this activity of course depends on how it is
distributed.  If the galaxies form as single entities then the bulk of
this activity will occur in $\sim 100$ massive gas-rich systems on a
timescale which is probably no more than 100-Myrs -- each must
therefore form stars at $\sim $200\,M$_\odot$\,yr$^{-1}$.  Locally such
high levels of star-formation activity are uniquely associated with
highly-obscured ultraluminous infrared galaxies (ULIRGs) such as Arp
220, which has L$_{FIR}\sim 3\times 10^{12}$L$_\odot$.  Dusty, gas-rich
galaxies of this luminosity would be identifiable via their strong,
redshifted far-infrared emission in the submillimetre (submm) waveband.
Alternatively, if the massive galaxies form through the more extended
merger of many subclumps then the individual star-formation events will
be much less intense, 10's M$_\odot$\,yr$^{-1}$.  Thus one simple route
to test the formation mechanism of massive galaxies is to search for
populations of strongly-clustered and gas-rich, luminous submm galaxies
(SMGs) at $z>2$--3, using submm cameras such as SCUBA on the JCMT
(Holland et al.\ 1999) or MAMBO on the IRAM 30-m.

\section{Locating SMGs and estimating their redshifts}

A population of apparently luminous submm sources has been identified
in deep maps with SCUBA and MAMBO, with a surface density of 0.25 per
sq.\ arcmin brighter than 5\,mJy at 850$\mu$m.  Moreover, gravitational
lens-augmented submm surveys show that most of the extragalactic
background in the submm arises in sources with apparent fluxes in the
decade between 5\,mJy and $\sim 0.5$\,mJy (e.g.\ Smail et al.\ 2002;
Cowie et al.\ 2002).  However, the book-keeping exercise of resolving
the submm background misses the critical issue --- {\it What is the
nature of the bright sources making up the background and how do they
relate to the formation of massive galaxies?}

Measuring the redshift distribution of SMGs has been a crucial aspect
of our interpretation of this population. In particular, with a
sufficiently accurate redshift for an SMG it is possible to search for
molecular CO emission which would confirm the masses and gas content of
these systems.  The measurement of a spectroscopic redshift for an SMG
came within weeks of the first deep SCUBA map, with the identification
of SMM\,J02399$-$0135 as a ULIRG Seyfert-2/BAL-QSO at $z=2.80$ (Ivison et
al.\ 1998). Subsequent studies of larger samples of SMGs took a variety
of approaches to try to identify counterparts in the optical and
near-infrared and then attempt to estimate the likely redshifts
photometrically or spectroscopically (Hughes et al.\ 1998; Smail et
al.\ 1998; Barger et al.\ 1999; Lilly et al.\ 1999; Frayer et al.\
2003).  The complication in identifying optical/near-infrared
counterparts to SMGs results from the large submm error-boxes (due to
the poor signal-to-noise of the detections and the large beam of SCUBA)
which may contain many plausible counterparts (Fig.~1).  As expected
for a young field, a fair proportion of the identifications initially
suggested proved incorrect, compare Smail et al.\ (2002) and Frayer et
al.\ (2003) with Smail et al.\ (1998) or Dunlop et al.\ (2003) with
Hughes et al.\ (1998), which at least demonstrates that we're making
progress.  What we learned was that there is no easy method to identify
an SMG from its optical magnitude or colours -- for example their
I-band counterparts span a range of 10,000$\times$ in brightness and
while they are on-average redder than the typical field galaxy, this
may not be a sufficiently unique feature to unambiguously locate the
submm source (rather than a companion).

In contrast, the radio waveband provides a relatively clean route to
accurately locate the position of the submm source.  This arises from
the tight correlation between the far-infrared emission from dust
heated by young, massive stars and the synchrotron emission in the
radio waveband, which has its origin in electrons accelerated in
supernovae remanents from the same young, massive stars.  This close
relationship between the far-infrared and radio emission suggests that
many SMGs should be detectable at 1.4\,GHz, and indeed 60--70\% of the
$>$5-mJy SMG population appear to have radio counterparts brighter than
about 30$\mu$Jy (Fig.~1; Chapman et al.\ 2002; Ivison et al.\ 2002)
[We'll discuss the missing third of the population in more detail
below].  The astrometric precision and resolution of 1.4-GHz VLA maps
taken in A/B configurations thus provides sub-arcsecond error-boxes for
any radio counterparts to SMGs, substantially better than possible from
the raw SCUBA (or MAMBO) maps.

As well as locating the SMGs, the radio waveband promised the
opportunity of estimating their redshifts, without any reference to
their optical/near-infrared counterparts, based on the form of the
spectral energy distribution (SED) in the submm/radio wavebands
(Carilli \& Yun 1999; Aretxaga et al.\ 2002; Efstathiou \&
Rowan-Robinson 2003).  Unfortunately, the precision of these estimates
is low and concerns have been voiced about the degeneracy between the
source redshift and its effective dust temperature (Blain 1999).  In
particular, these uncertainites mean the estimates are too crude to
reliably target the sources for CO follow-up.  In addition the
redshift/temperature degeneracy means it is impossible to
simultaneously estimate the redshifts {\it and} bolometric luminosities
of these sources from their submm/radio colours and hence derive star
formation densities for the SMG population.

The low precision and systematic degeneracies in the submm/radio
photometric redshifts mean we still need a method for determining
accurate and reliable redshifts for a large sample of SMGs. Until
sensitive, broad-band CO millimetre spectrometers are in use we are
forced back to measuring redshifts in the optical/near-infrared
wavebands.  Such deep spectroscopic observations on a case-by-case
basis had yielded redshifts for only about seven robustly-identified
SMGs, at $z$=0.66--3.35, by the beginning of 2002.  To make substantial
progress in understanding the properties of SMGs we would need a sample
an order of magnitude larger and hence we would need a more efficient
technique.  Here we are again aided by information from the radio, but
rather than fluxes it is the precise positions which are the key.

%
%
\begin{figure}
\centerline{\psfig{file=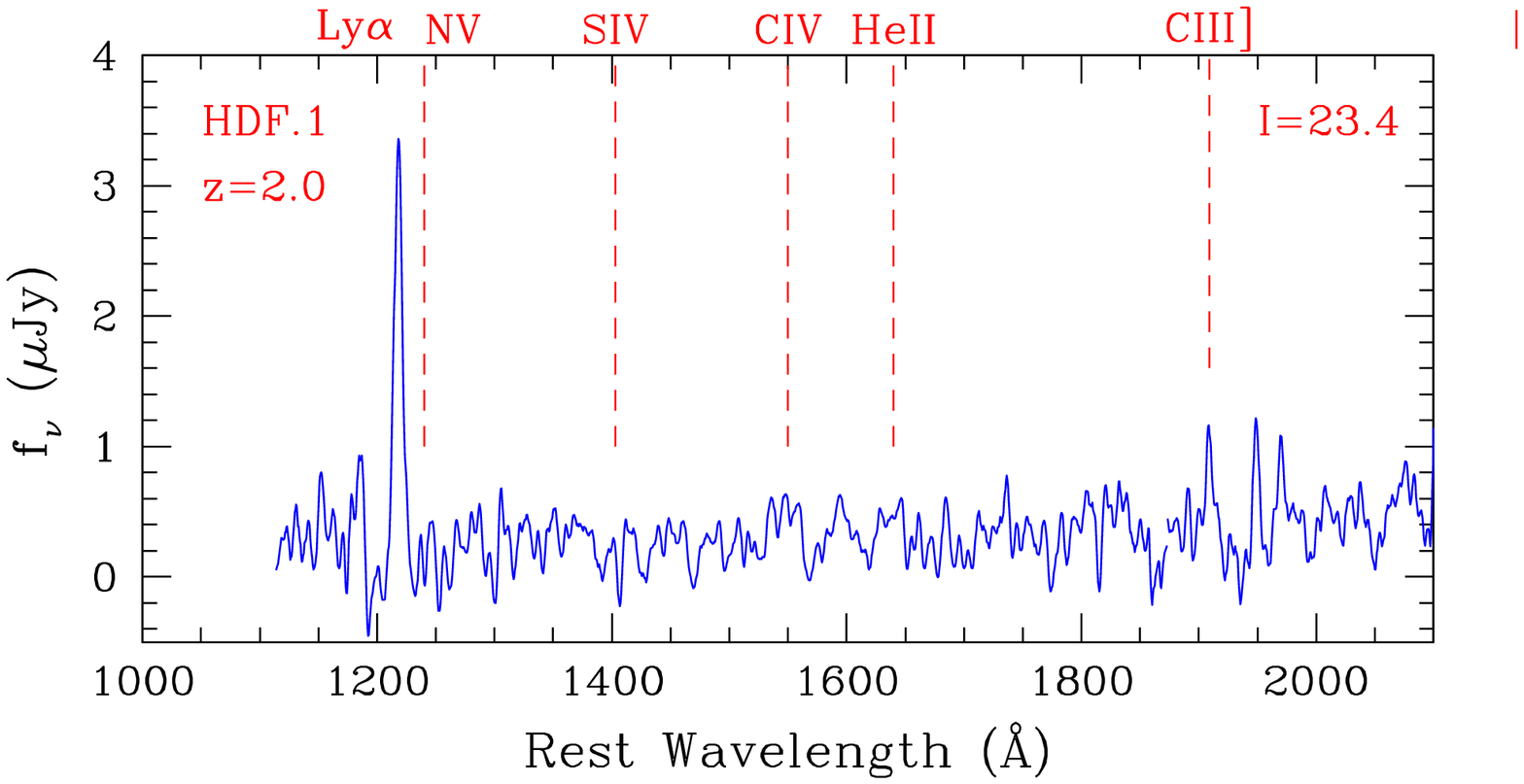,width=2.8in,angle=0}
\psfig{file=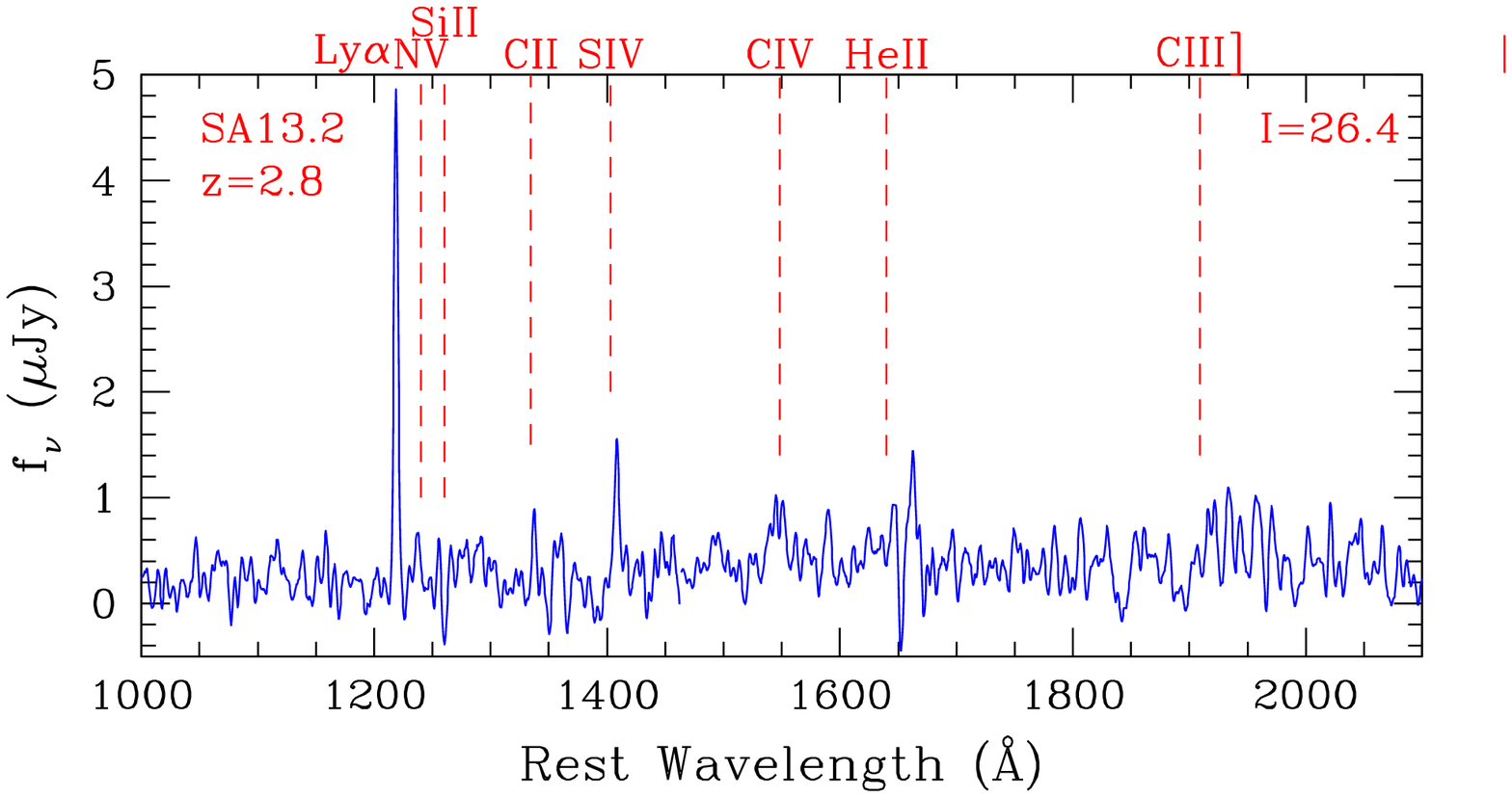,width=2.8in,angle=0}}
\caption{\small
Two examples of the spectra of SMGs from the Keck/LRIS-B survey of
Chapman et al.\ (2003 \& in prep.).  These demonstrate the strength of the
emission lines (in particular Ly-$\alpha$) in these optically-faint
galaxies, as well as the need for high efficiency in the UV/blue in the
spectrograph to detect such lines at $\ls 3600$\AA.
}
\end{figure}

\section{Measuring redshifts for SMGs}

For the two-thirds of SMGs with $\mu$Jy radio counterparts we have
positions precise enough to place a spectrograph slit on the submm
source (even if we can't see an optical/near-infrared counterpart).  By
exploiting both wide-field submm surveys and targeted submm
observations of radio sources it is possible to build-up high enough
surface densities of SMGs that a reasonable multiplex gain ($\sim
6\times$) can be obtained within the typical field-of-view of a
multi-object spectrograph on an 10-m class telescope, thus justifying
exposure times of a few hours per field.  Using this approach Chapman
et al.\ (2003 \& in prep.) have measured redshifts for 66 SMGs (Fig.~2 \& 3).
The two key elements of this survey, in addition to the radio-derived
positions, are: the use of a highly-efficient, blue-optimised
spectrograph: LRIS-B on Keck; and a cavalier disregard for the apparent
magnitude of any optical counterpart.

The high through-put achieved with LRIS-B right down to the atmospheric
cut-off at $\sim 3100$\AA, represents a substantial gain over any other
spectrographs on 8/10-m class telescopes.  Its total through-put at
4000\AA\ is $\sim 40$\% (including the telescope), compared to $\sim
15$--20\% for a single-beam spectrograph such as FORS or GMOS.  More
importantly, the through-put of LRIS-B drops only a third as fast as
GMOS or FORS going from 4000\AA\ to 3200\AA.  Thus detecting the third
of the SMG population at $z=1.8$--2.3 which have Ly-$\alpha$ at
$<4000$\AA, is an order-of-magnitude easier with LRIS-B (Fig.~2).

As with all redshift surveys of multiwavelength samples, the issue of
spectroscopic completeness is complicated by the huge range in optical
properties of SMGs.  Nevetheless, we note that these sources, while
sometimes faint in a particular optical passband, can still have strong
emission lines at other wavelengths. Hence, the incompleteness is
likely to be determined by the absence of strong lines in the observed
wavelength region, rather than the exact continuum signal-to-noise
achieved (which will usually be low).  Moreover, as only a small
fraction of their luminosity comes out in the UV/optical, $\ls 10$\%,
the continuum properties in these wavebands are unlikely to be critical
for our basic understanding of the characteristics of SMGs.  Therefore
we concluded that trying to produce a ``complete'' magnitude-limited
redshift survey of SMGs will prove difficult (and is irrelevant
anyway).

We have therefore targeted a sample of radio-detected SMGs with
I=22--27 in seven fields (HDF, Lockman Hole, ELAIS-N2, SA13, SA22, CFRS
3hr and 14hr) with no regard to their optical magnitudes.  The submm
data comes from SCUBA mapping surveys in these areas, supplemented by
photometry-mode observations of samples of optically-faint radio
galaxies.  The submm sources have 850-$\mu$m fluxes of 3--20 mJy, with
a median of $5.7\pm 1.8$\,mJy.  In addition, we include a number of
other sources in our masks: candidate counterparts or companions to
SMGs without radio detections, identified through their extreme colours
(e.g.\ EROs) or simply their optical properties (the latter are
somewhat analogous to those targeted by Barger et al.\ 1999). These
sources could be SMGs at somewhat higher redshifts than the
radio-identified population (as the radio fluxes of SMGs at $z>3$ fall
below our detection limit) or galaxies with colder characteristic dust
temperatures lying at similar redshifts to the radio-identified
population.  We also include optically-faint $\mu$Jy radio galaxies
without SCUBA observations which could be subsequently targeted with
SCUBA. Finally, we add in samples of UV-selected populations at $z \sim
2$--3.5 (Steidel et al.\ 2003 \& in prep.).  These are included to allow us
to relate the clustering and environments of these various classes of
high-redshift galaxies.

%
%
\begin{figure}
\centerline{\psfig{file=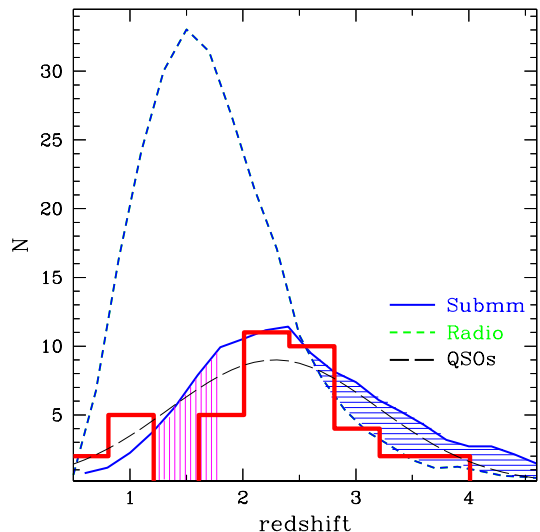,width=3.0in,angle=0}}
\caption{\small
The redshift distribution from the Keck SMG survey (Chapman et al.\
2003 \& in prep.).  The histogram shows the distribution of radio-identified
SMGs, which has a median redshift of $<\!z\! >=2.3\pm 0.4$ (remarkably
similar to the behaviour of optically-bright QSOs, shown by the
long-dashed line).  We describe the selection function of the survey
using an evolutionary model for the submm population convolved with the
radio selection function. The model accounts for the dust properties of
local luminous infrared galaxies and has been tuned to fit the faint
submm counts.  We plot the predicted redshift distributions for SMGs
with S$_{850\,\mu m}>5$\,mJy (solid line) and radio sources with
S$_{1.4\,\rm GHz}>30$\,$\mu$Jy (dashed line).  We expect to miss
sources at $z\gs 3$ lying between the submm and radio curves (horizontal
shading) due to the need for a radio detection to pinpoint the SMGs.
The apparent deficit between model and observations at $z\sim1.5$
(vertical shading) probably reflects the difficulty in obtaining spectra in
the spectroscopic desert.
}
\end{figure}

Exposure times ranged between 2--6\,hrs per mask, with the equivalent
of 10 clear nights used for the programme.  The spectroscopic
identification rate is high for such faint galaxies, $\sim 70$\%,
underlining the frequent occurrence of strong emission lines in many
systems, including Ly-$\alpha$, a surprising line to detect in these
dusty galaxies (Fig.~2).

The median redshift of the population is $<\! z\! >=2.3$ with a
quartile range of $z$=1.9--2.6 (Fig.~3) and a space-density within the
quartile range of $\sim 10^{-5}$\,Mpc$^{-1}$.  This confirms that most
SMGs are luminous sources at sufficiently high redshifts that they
could be the precursors of the massive galaxies identified at $z\sim
1$--2, as well as the progenitors of the old, massive elliptical
galaxies seen at the present-day.  If the submm-luminous phase of the
SMG's evolution lasts for $\sim 100$\,Myrs (Smail et al.\ 2003a), this
is roughly 10\% of the cosmic time spanned by the quartile redshift
range of the population.  Hence to estimate the density of the
present-day counterparts to this population we need to correct the
apparent space density for the visibility of the bright SMGs at these
redshifts by a factor of 10$\times$.  The descendents of the SMGs
will therefore have a space density of $\sim 10^{-4}$\,Mpc$^{-3}$,
comparable to $\geq $\,L$^\ast$ ellipticals.  A comparison of the SMG
redshift distribution and that from the ``classical'' LBG photometric
selection (Steidel et al.\ 2003) shows only a modest overlap, $\ls 30$\%,
which might help explain the weak correlation between the two
populations  (Chapman et al.\ 2000; Webb et al.\ 2003).

The incompleteness in our radio-detected SMG survey, compared to a
putative complete, purely submm-selected redshift survey has two
components: a) the $\sim 30$\% of radio-detected SMGs where we fail to
measure a redshift; b) the $\sim 35$\% of the SMG population which lack
radio counterparts brighter than $\sim 30\mu$Jy.  The former can be
attributed to galaxies falling in the spectroscopic desert at $z\sim
1.5$ and SMGs lacking strong emission lines (at similar redshifts to
those we detect).  While the latter may be either due to: i) sources
with somewhat colder dust at similar redshifts to the bulk of the
population; ii) a separate SMG population at much higher redshifts, $z>
3$; iii) spurious submm detections.  As we show in Fig.~3, the
incompleteness due to radio non-identifications is consistent with that
expected from the distribution of dust temperatures in the local
luminous infrared galaxies (Chapman et al.\ 2003).  This would
place the missing population in the same volume as that probed by the
radio-identified sources, which is supported by the redshifts of many
of the tentative ERO and non-radio candidate SMG identifications.
Alternatively, the lack of radio, X-ray and in some cases
millimetre-band, counterparts to a modest fraction of SCUBA sources
may simply point to some of these low significance detections being
spurious (Alexander 2003; Greve et al.\ 2003).

\section{Masses of SMGs}

Armed with (relatively) precise redshifts for a substantial sample of
luminous SMGs we can now directly tackle the issue of the dynamics, gas
masses and fractions of this population to test whether these are
consistent with those required for the monolithic progenitors of
massive galaxies.  To this end we are undertaking a long-term survey of
the molecular CO emission from this population using the IRAM Plateau
de Bure interferometer (PI's Reinhard Genzel \& Rob Ivison; Neri et
al.\ 2003).  This builds upon the earlier CO follow-up of SMG in the
SCUBA Lens Survey (Frayer et al.\ 1998, 1999), and to date 10 SMGs are
detected in CO, at $z=1.06$--3.35 of which data for 5 have so far been
published (Neri et al.\ 2003; Greve et al.\ in prep.).  The mean 850$\mu$m
flux for these sources is 6.4-mJy, corresponding to a far-infrared
luminosity of $12\times 10^{12}$\,L$_\odot$ and a star formation rate
of $\sim 1000$\,M$_\odot$\,yr$^{-1}$ for stars more massive than
5\,M$_\odot$.  Their mean restframe V-band luminosity is $2\times
10^{11}$L$_\odot$ (based on their K-band magnitudes and corrected for
lensing where appropriate) suggesting minimum starburst masses of $\gs
1\times 10^{10}$M$_\odot$, assuming mass-to-light ratios typical of
$\sim 10^7$\,Myr-old starbursts and a Salpeter IMF.  The FWHM of the CO
lines of the ensemble is $500\pm 110$\,km\,s$^{-1}$ (this compares to
260\,km\,s$^{-1}$ for the FWHM of H$\alpha$ emission lines in LBGs, Erb
et al.\ 2003), with the emission typically spatially-unresolved ($\ls
10$\,kpc) giving dynamical mass of $\ls 1 \times 10^{11}$\,M$_\odot$
assuming the systems are merging (c.f.\ Genzel et al.\ 2003; Neri et
al.\ 2003).  The H$_2$ masses are similarly high, $1\times
10^{11}$M$_\odot$ (assuming low--metallicity gas and hence
$X$(CO/H$_2$)=5, these gas masses would drop by a factor of $5\times$
if we used $X$(CO/H$_2$) for local ULIRGs) implying that the central
regions are likely to be baryon-dominated.  These gas reservoirs would
support the observed star formation rates for 50\,Myrs (assuming half
is lost through outflows) and produce an additional $\sim$\,M$^\ast$'s
worth of stars in that time.  Truncating or tilting the IMF to provide
a larger fraction of high-mass stars will potentially prolong the
active phase (and decrease the visibility correction to the estimated
space density).  Such a tilt would also explain the large metal yield
required to produce the iron abundance in the intracluster medium.

The availability of precise redshifts for a large sample of SMGs
has allowed us to confirm that these are gas-rich, massive galaxies
with individual characteristics consistent with the progenitors of $\gs
$\,L$^\ast$ massive galaxies seen at the present-day.  Such galaxies
are expected to be strongly clustered and we can also test this with
our sample.

\section{Clustering of SMGs}

Trying to measure the clustering strength of SMGs from their
distribution on the sky is one situation where the small number
statistics in current surveys and the very broad redshift selection
function of this population conspire to make the estimates very
uncertain (Scott et al.\ 2002; Borys et al.\ 2003: Webb et al.\ 2003).
Even the largest area submm surveys correspond to 1:300 aspect ratio
pencil-beams, with widths $\ls 10$\,Mpc and lengths of 3\,Gpc.  Hence,
redshifts for SMGs are again very helpful for making progress.

Using our spectroscopic survey we can identify SMGs which lie close to
each other in redshift. We therefore search for small-velocity
separation associations ($\leq 1200$\,km\,s$^{-1}$) within each field
and find four close-pairs of SMGs, a triplet and quintet all with
projected separations of $\ls 10$\,Mpc on the sky (Fig.~4).  The strong
clustering of the SMG population is underlined by the fact that we can
reliably detect the clustering signal in a sample of only 66 SMGs
spread across seven distinct fields.  The likelihood of finding these
systems by random chance is $<$0.01\% and using Monte Carlo simulations
we can constrain the correlation length of the sample to be r$_o\sim
8\pm3$h$^{-1}$\,Mpc.  This suggests that SMGs represent one of the most
strongly clustered populations of high-redshift sources known (Fig.~4).
The simple interpretation of this clustering is that SMGs are a
highly-biased population residing in massive halos, $\gs
10^{13}$M$_\odot$.  Their clustering strength (although uncertain) is
consistent with them being the precursors of the passive ERO population
at $z\sim 1$ and perhaps evolving into the massive gE's seen in
clusters at the present-day (Fig.~4).

%
%
\begin{figure}
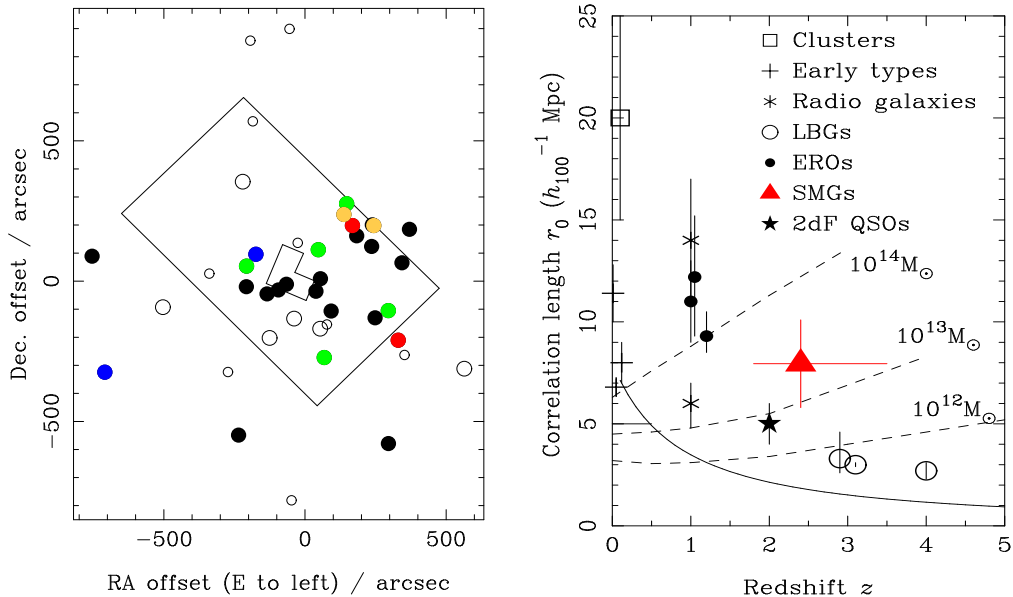

\centerline{\psfig{file=smail_fig4a.ps,width=2.5in,angle=270} ~~~
\psfig{file=smail_fig4b.ps,width=2.5in,height=3.1in,angle=270}}
\caption{\small [left] The distribution of radio-identified SMGs around
the Hubble Deep Field (filled points), these are colour-coded on their
redshifts, SMGs within 1200\,km\,s$^{-1}$ of each other share the same
colour (the black points are not paired).  There are three pairs and a
quintet of SMGs all lying within narrow velocity ranges and projected
separations of $\ls 10$\,Mpc.  The two overlays show the coverage of the
HDF WFPC2 and GOODS fields and the open symbols are
spectroscopically-unidentified SMGs. [right] A comparison of the
correlation lengths of different populations of high-redshift sources
(adapted from Overzier et al.\ 2003).  The dashed lines show the
clustering of different fixed-mass halos as a function of redshift,
while the solid line illustrates the evolution of the clustering
strength of a population.  The $\gs5$\,mJy SMGs at $z\sim 2$--3 appear
to be more strongly clustered than either the LBG or QSO populations at
this epoch (Blain et al.\ 2003).  This is consistent with SMGs 
typically residing in  more massive halos, $\gs 10^{13}$M$_\odot$.
}
\end{figure}

There is also evidence for clustering of the SMG population with other
classes of high-redshift galaxies.  For example, SMGs are found in the
over-dense regions around high-redshift radio galaxies, which are also
traced by LBGs, {\it Chandra} X-ray sources and Ly-$\alpha$ galaxies
(e.g.\ Stevens et al.\ 2003; Smail et al.\ 2003b).  Less obscured
galaxies are frequently found within $\ls 1$\,Mpc of SMG's, such as the
$z=2.5$ companion to the multiply-imaged SMG behind A\,2218 (Kneib
et al.\ in prep.)  or the Ly-$\alpha$ emitter at $z=2.8$ close to
SMM\,J02399$-$0135 behind A\,370 (Santos et al.\ 2003).  Equally the
multiplets of SMGs appear to pin-point the redshifts of strong
overdensities of UV-selected systems -- for example the triplet of SMGs
in the SA22 field lies in the well-studied $z=3.09$ ``spike'' of
LBGs/Ly-$\alpha$ sources (Steidel et al.\ 2000).  This provides
circumstantial support for the claim that SMGs reside in the highest
density regions at $z=2$--3, which evolve into the cores of clusters at
the present-day.

\section{Discussion and Conclusions}

To summarise, we have exploited the broad overlap between bright submm
sources ($>5$\,mJy) and $\mu$Jy radio sources ($>30\mu$Jy) to precisely
locate the sources of the submm emission.  By targeting these with
LRIS-B/Keck we have measured redshifts for at least half of the bright
SMG population, confirming that these are high redshift, ultraluminous
infrared galaxies at $z\sim 2.3$.  The space density of these SMGs is
$\sim 1000\times$ higher than similar luminosity galaxies at the
present-day and once the likely timescale for their visibility is taken
into consideration they have a space density comparable to $\gs
$\,L$^\ast$ ellipticals at $z=0$.  

The precise redshifts for luminous SMGs enable us to search for CO
emission from these systems using millimetre interferometers.  These
observations probe the gas mass and dynamics in the inner regions of
the SMGs, showing that these are both massive and extremely gas-rich
systems: median dynamical and gas masses of $\sim 10^{11}$\,M$_\odot$.
The typical bright SMG in our survey hosts a reservoir of gas which is
sufficient to produce the stellar mass of an $>$L$^\ast$ galaxy on a
timescale $\ls 100$\,Myrs (in addition to their substantial existing
stellar populations).  These characteristics are precisely those
expected for the prompt formation of massive galaxies through a rapid
starburst at high redshifts.  The connection between the high-redshift
luminous SMG population, passive EROs at $z=1$ and massive galaxies at
$z=0$ is further strengthened by the detection of strong clustering in
the SMG population.  This analysis also relies on the precise redshift
measurements available from our Keck survey and shows that the SMGs at
$z\sim 2$--3 may be more strongly clustered than either optical QSOs or
LBGs at these epochs (although see Scannapieco \& Thacker 2003).  If
the SMG halos are really this strongly clustered then it makes it
difficult to see how these systems could evolve into typical LBGs. We
note that this difference in the clustering strength of bright SMGs and
typical LBGs is in the same sense as the factor of $\gs 4\times$ mass
difference implied by their respective CO and H$\alpha$ line widths.
However, claims of stronger clustering in more luminous LBGs may
provide a link between these two populations, as will on-going studies
of the relative distribution of SMGs and the less luminous, but more
common the UV-selected population at $z\sim 2$--3.

\section*{Acknowledgements}
We thank our collaborators on the Plateau de Bure survey (Frank
Bertoldi, Pierre Cox, Reinhard Genzel, Thomas Greve, Roberto Neri,
Alain Omont and Linda Tacconi).  IRS acknowledges support from the Royal
Society and Leverhulme Trust. AWB acknowledges support through NSF
grant AST-0205937.

\end{document}